%% file: Notices_ArXiv.tex
\documentclass{article}
\usepackage{multicol}
\usepackage[margin=.5in]{geometry}
\usepackage{amsmath, amssymb, amsfonts, amsthm}
\usepackage{bm}
\usepackage{subfigure}
\usepackage{float}
\usepackage{graphicx}
\usepackage{xcolor}
\usepackage{transparent}
\usepackage{mathabx}
\usepackage{mathtools}
\usepackage{multirow}
\usepackage{setspace}
\usepackage[all]{xy}
\usepackage{epsfig}
\usepackage{slashed}
\usepackage{abstract}

\allowdisplaybreaks
\newcommand{\ubar}{\underline}

\newtheorem{theorem}{Theorem}
\newtheorem{definition}{Definition}

\DeclareMathOperator{\II}{II}

\author{Hubert L. Bray and Henri P. Roesch}
\title{Null Geometry and the Penrose Conjecture}
\date{}
\begin{document}
\maketitle
\begin{abstract}
\noindent ABSTRACT. In this paper, we survey recent progress on the Null Penrose Conjecture, including a proof of the conjecture for smooth null cones that are foliated by doubly convex spheres.
\end{abstract}
  \begin{multicols}{2}
 \noindent Sir Roger Penrose argued in 1973 \cite{penrose1973naked} that the total mass of a spacetime containing black hole horizons with combined total area $|\Sigma|$ should be at least $\sqrt{|\Sigma|/16\pi}$. On the one hand, this conjecture is important for physics and our understanding of black holes. On the other hand, Penrose's physical arguments lead to a fascinating conjecture about the geometry of hypersurfaces in spacetimes. For spacelike (Riemannian) slices with zero second fundamental form the conjecture is known as the \textit{Riemannian Penrose Inequality} and was first proved by Huisken-Ilmanen \cite{huisken2001inverse} (for one black hole) and then by the first author \cite{bray2001proof}, using two different geometric flow techniques. This paper concerns a formulation of the conjecture for certain null hypersurfaces in spacetimes called the \textit{Null Penrose Conjecture (NPC)}. Over the last ten years, there has been a great deal of progress \cite{alexakis2015penrose, christodouloumathematical2,ludvigsen1983inequality, mars2015asymptotic, mars2016penrose, sauter2008foliations} on the NPC, culminating in a proof of the conjecture in a fair amount of generality for smooth null cones \cite{roesch2016proof}. One surprising fact is that these null hypersurfaces, under physically inspired curvature conditions on the spacetime, have monotonic quantities, including cross sectional area \cite{1973ellis}, notions of energy \cite{ludvigsen1983inequality, mars2015asymptotic, mars2016penrose, sauter2008foliations}, and, as we'll see with Theorem 1 below, a new notion of mass \cite{roesch2016proof}.\\\\
\noindent\textbf{The Null Geometry of Light}\\
\noindent The theory of General Relativity emerges from Albert Einstein's beautiful idea that matter in a physical system curves the intertwining fabric of both space and time. A \textit {spacetime} is a four dimensional manifold with a metric of signature (3,1), meaning that the metric on each tangent plane is isometric to the Minkowski spacetime $\mathbb{R}^4_1:=(\mathbb{R}^4, - dt^2 + dx^2 + dy^2 + dz^2)$. The minus sign in the metric implies the existence of null vectors, vectors with zero length (such as (1,1,0,0)), even though the vectors themselves are not zero. A null hypersurface is a codimension one submanifold of a spacetime whose three dimensional tangent planes are null in one dimension (and hence positive definite in the two other dimensions). For example, if we let $r = \sqrt{x^2 + y^2 + z^2}$ in the Minkowski spacetime, any translation of the downward cone $\Lambda:=\{t = -r\}$ as depicted in Figure 1, is a null hypersurface. \\
\indent Null geometry is counter intuitive in a number of ways.  Since one dimension has zero length, null hypersurfaces have zero volume. Furthermore, since the metric is not invertible, the Riemann curvature tensor of the null hypersurface is not well defined. Also, the vector which is perpendicular to a null hypersurface must also be tangent (hence null). This normal-tangent duality complicates the notion of what the second fundamental form of a null hypersurface should be defined to be (the classical tool for analyzing the ``shape" of substructures). 

\begin{figure}[H]
\centering
\def\svgwidth{\linewidth}
\fbox{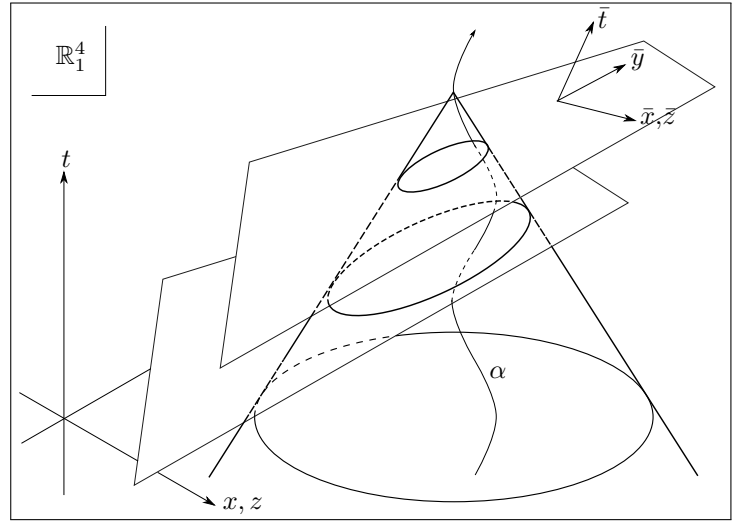}
\caption{Downward Cone of Minkowski, $\Lambda$}
\end{figure}

\noindent Fortunately, for a `conical' null hypersurface $\Omega\cong\mathbb{S}^2\times\mathbb{R}$ called a \textit{null cone}, where $\mathbb{S}^2$ accounts for the two positive dimensions, $\Omega$ can be studied vicariously through the geometry of its spherical cross-sections including their Gauss and Codazzi equations. For our sacrifice in intuition to this normal-tangent duality we do gain some advantages. A direct consequence of a normal vector $\ubar L$ being also tangent is that all curves along $\ubar L$ must be geodesic. Thus, a null hypersurface can be thought of as a collection of `light-rays' in the framework of General Relativity. Imagine standing on some 2-sphere $\Sigma_0$ in a spacetime, for example the surface of the Earth, and collecting all light rays `hitting the surface' at a particular point in time. 

\begin{figure}[H]
\centering
\def\svgwidth{\linewidth}
\fbox{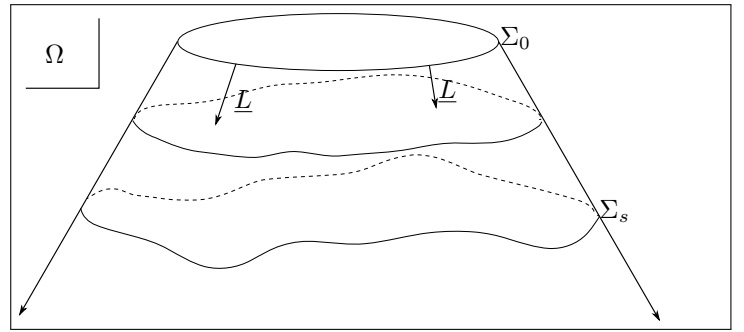}
\caption{Foliations of $\Omega$}
\end{figure}

\noindent The resulting set constructs (or recovers) a null cone $\Omega$ reducing the usually complicated system of PDEs associated to flows in spacetimes to an analysis of ODEs. From standard uniqueness of ODEs, any two normal null flows of surfaces off of $\Sigma_0$ must result in two foliations of the same null cone. Moreover, from standard existence of ODEs, wherever there's a smooth spacelike 2-sphere $\Sigma_0$ there will be a null cone off of it, at least in a neighborhood of the sphere.\\\\
\noindent\textbf{The Expanding Null Cone}\\
The thermodynamics of black holes rests upon Hawking'€™s area theorem \cite{1973ellis} which states that in spacetimes satisfying the Dominant Energy Condition (or DEC) the area of a cross section of a black hole event horizon is non-€"decreasing. Similarly in our context, we will briefly explore how the DEC, which is a local curvature constraint modelling non-negative energy, ensures that null cones reaching \textit{null infinity} can only be foliated by spheres of increasing area. To do so, we first need a quantity also needed to state our main result, Theorem 1.
\begin{definition}
\emph{For a spacelike 2-sphere $\Sigma$, the \textit{expansion} is given by 
$$\sigma:=\langle-\vec{H},\ubar L\rangle$$
where $\vec{H}$ is the \textit{mean curvature}, a normal vector measuring the mean extrinsic curvature of $\Sigma$.}
\end{definition}
For a null flow off of $\Sigma$ along $\ubar L$, $\sigma$ comes from $\frac{d}{ds}dA = \sigma dA$ where $dA$ is the area form (i.e. an `element of area') on $\Sigma$. Here the DEC comes into play, by way of the Raychaudhuri equation (see \cite{1973ellis}), ensuring $d\sigma/ds\leq0$ along all geodesics. Hence, the only way to get standard asymptotics at infinity, which by any reasonable notion should have expanding 2-spheres (i.e ``$\sigma(\infty)>0$"), necessarily restricts our choice of $\Sigma_0$ to have strictly positive expansion. In fact, $\sigma>0$ on all of $\Omega$ enforces that all foliations must have expanding area.\\
\indent With the null flow vector $\ubar L$ having zero length we also forfeit our intuitive notion of measuring the speed of a flow. However, with an expanding null cone $\Omega$, $\sigma$ offers a convenient replacement.\\
\indent Null cones that reach null infinity also have fundamental physical significance beyond the monotonicity of cross-sectional area. In isolated physical systems, such as a cluster of stars or a black hole, we expect the curvature induced by localized matter to settle ``far away" back to flat Minkowski spacetime. In our context, Mars and Soria \cite{mars2015asymptotic} introduced the notion of an \textit{asymptotically flat} null cone whereby the geometry of $\Omega$ approaches that of a downward cone of Minkowski at null infinity in a suitable sense. These asymptotically flat null cones allow us the ability to measure the total energy and mass of a system which we'll need in order to state the NPC.\\\\ 
\noindent\textbf{Total Energy and Our Main Example}\\
In 1968, Hawking \cite{hawking1968gravitational} published a new mechanism aimed at capturing the amount of energy in a given region using the curvature of its boundary $\Sigma$. 
\begin{definition}
\emph{The \textit{Hawking Energy} is given by}
\begin{equation}
E_H(\Sigma) = \sqrt{\frac{|\Sigma|}{16\pi}}\Big(1-\frac{1}{16\pi}\int_\Sigma\langle\vec{H},\vec{H}\rangle dA\Big).
\end{equation}
\end{definition}
\indent For example, for any cross-section of the downward cone $\Sigma\hookrightarrow\Lambda:=\{t=-r\}$ in Minkowski the Gauss equation identifies a beautifully simple relationship between the intrinsic and extrinsic curvature, $\mathcal{K}=\frac14\langle\vec{H},\vec{H}\rangle$ (see \cite{roesch2016proof}), where $\mathcal{K}$ is the Gauss curvature of $\Sigma$. From the Gauss-Bonnet Theorem we therefore conclude that $0=\int\mathcal{K}-\frac14\langle\vec{H},\vec{H}\rangle dA = 4\pi E_H/\sqrt{|\Sigma|/16\pi}$. Thus, all cross-sections - no matter how squiggly - envelop matter content of vanishing energy, as expected of a flat vacuum.\\
\indent For another, and our main example of a null cone, we go to the one parameter family of Schwarzschild spacetimes characterized by the metric
$$-(1-\frac{2M}{r})dt^2+\frac{dr^2}{1-\frac{2M}{r}}+r^2(d\vartheta^2+(\sin\vartheta)^2d\varphi^2).$$
These spacetimes model an isolated black hole with the parameter $M$ representing total mass. Note that the coordinate $r$ has been chosen so that each sphere of fixed $(t,r)$ is a round sphere of area $4\pi r^2$. Also, when $M=0$ we recover exactly the Minkowski metric in spherical coordinates. The reader may have noticed the singularities $r=0$ and $r=2M$. The singularity at $r=0$ is a curvature singularity called the \textit{black hole singularity} giving rise to the isolated black hole. We see this black hole is isolated from the fact that the metric approaches the Minkowski metric for large values of $r$. On the other hand, the singularity at $r=2M$ is superficial, and can be removed with a change of coordinates. To show this we introduce the ingoing null coordinate $v$, whereby $dv=dt+\frac{dr}{1-\frac{2M}{r}}$, giving the metric in \textit{ingoing Eddington-Finkelstein} coordinates
\begin{equation}
-(1-\frac{2M}{r})dv^2+2dvdr+r^2(d\vartheta^2+(\sin\vartheta)^2d\varphi^2).
\end{equation}
\indent In the transition from $M=0$ to $M>0$ the downward cones of Minkowski transition to their spherically symmetric counterparts in Schwarzschild, referred to as the \textit{standard null cones}. In Minkowski, $\Lambda=\{v = 0\}$ for $v=t+r$, from (2) we see the analogous three dimensional slice $\Omega_S:=\{v=v_0\}$ (i.e. $dv=0$) inherits the metric $r^2(d\vartheta^2+(\sin\vartheta)^2d\varphi^2)$ assigning positive lengths only for vectors along the two spherical coordinates $(\vartheta,\varphi)$, not $r$. These coordinates $(r,\vartheta,\varphi)\in\mathbb{R}\times\mathbb{S}^2$ identify points on the standard null cone $\Omega_S$. 

\begin{figure}[H]
\centering
\def\svgwidth{\linewidth}
\fbox{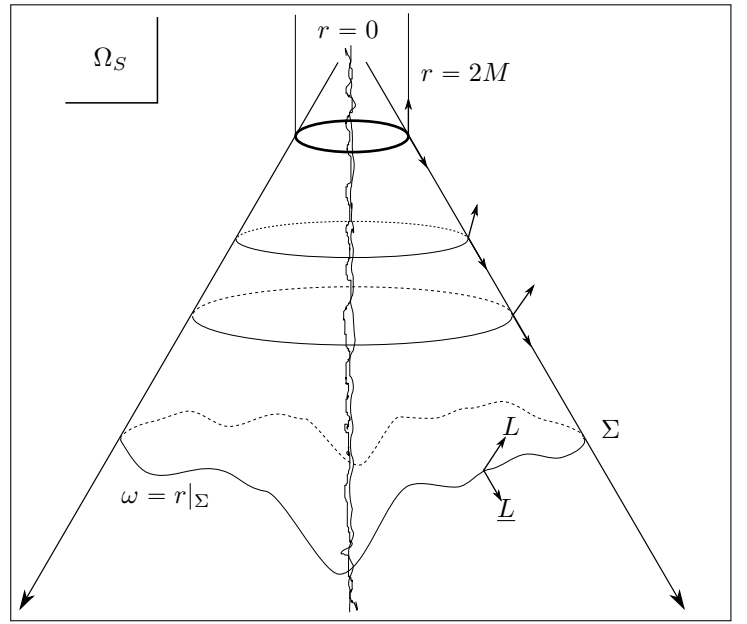}
\caption{The Standard Null Cone, $\Omega_S$}
\end{figure}

The $r$-coordinate curves are exactly the geodesics that rule $\Omega_S$, allowing us to identify any cross-section by simply specifying $r$ as a function on $\mathbb{S}^2$. Similar to Minkowski, the Gauss equation once again simplifies to an intriguingly simple expression for a cross-section $\Sigma:=\{r=\omega(\vartheta,\varphi)\}\hookrightarrow\Omega_S$ (\cite{roesch2016proof}) 
\begin{equation}
\mathcal{K} - \frac14\langle\vec{H},\vec{H}\rangle = \frac{2M}{\omega^3}.
\end{equation}
$\Sigma$ also inherits the simple metric $\gamma = \omega^2\mathring\gamma$ from $\Omega_S$, where $\mathring\gamma$ is the standard round metric on a sphere. So from Gauss-Bonnet and (3), we conclude that $E_H(\Sigma)=M\sqrt{\int\omega^2\frac{dS}{4\pi}}\int \frac{1}{\omega}\frac{dS}{4\pi}$ for $dS$ the area form on a round sphere. Some fascinating observations follow from Jensen's inequality. We deduce that $E_H\geq M$ which the reader may recognize as the special relativistic notion that the energy of a particle is always bounded below by its mass. Jensen's inequality also ensures that equality is reached only if $\omega = r_0$, corresponding to the $t$-slice intersections with $\Omega_S$ (see Figure 4). Moreover, one can show (\cite{mars2015asymptotic}) that $\Sigma$ is a round sphere if and only if $\omega(\vartheta,\varphi) = r_0\frac{\sqrt{1-|\vec{v}|^2}}{1-\vec{v}\cdot\vec{n}(\vartheta,\varphi)}$, for some $r_0$, $\vec{v}$ inside the unit ball $\mathring{B}^3\subset\mathbb{R}^3$, and $\vec{n}(\vartheta,\varphi)$ the unit position vector, giving the energy $E_H(r_0,\vec{v}) = \frac{M}{\sqrt{1-|\vec{v}|^2}}$. This is precisely the observed energy of a particle of mass $M$ traveling at velocity $\vec{v}$ relative to its observer (with the speed of light set to $c=1$).\\
\indent Using Schwarzschild as an example we can also motivate the notion of total energy and mass for an asymptotically flat null cone $\Omega$. We start by bringing to the attention of the reader that the intrinsic geometry of $\Omega_S$ is identical to that of the downward cone in Minkowski (\cite{sauter2008foliations}). This is evident from (2), since the only component `giving rise' to the Schwarzschild geometry (beyond Minkowski) is $dv^2$, which vanishes for the induced metric on $\Omega_S$. So instead, we may actually account for all round spheres of $\Omega_S$ by intersecting the downward cone of Minkowski by Euclidean hyperplanes (see Figure 1). On the one hand, any family of parallel hyperplanes in Minkowski are given as fixed time slices inside a reference frame (coordinates $(\bar{t},\bar{x},\bar{y},\bar{z})$) of an observer traveling at velocity $\vec{v}$. On the other hand, the ambient geometry of Schwarzschild settles to that of Minkowski, inheriting such characteristics asymptotically. Therefore, aided by Figure 4, we imagine that an asymptotically round foliation of $\Omega_S$ is induced by an asymptotically Euclidean slicing of the spacetime (for which $E_H$ has a verified correlation to total energy \cite{huisken2001inverse}). We conclude that an observer `at infinity' approximates our black hole to a particle (similarly to $\alpha$ of Figure 1) of total energy $\frac{M}{\sqrt{1-|\vec{v}|^2}}$. In the general setting, considering an asymptotically flat null cone $\Omega$, it follows that $E_H$ approaches a measure of total energy along an asymptotically round foliation called a \textit{Bondi Energy}, $E_B$ (see \cite{mars2016penrose}). 
\begin{definition}
\emph{For an asymptotically flat null cone $\Omega$, minimizing over all Bondi Energies $E_B$ yields the \textit{Bondi Mass} $m_B$ (\cite{mars2016penrose}).}
\end{definition}
\noindent Returning to Schwarzschild, we verify that the standard null cone has Bondi Mass $m_B(\Omega_S) = \displaystyle{\inf_{|\vec{v}|<1}}\frac{M}{\sqrt{1-|\vec{v}|^2}} = M$.\\\\
\textbf{Black Holes and the Null Penrose Conjecture}\\
Upon further inspection of (2) the reader may have noticed yet another null cone given by the slice $\mathcal{H}:=\{r=2M\}$ (see Figure 2). With induced metric $4M^2(d\vartheta^2(\sin\vartheta)^2d\varphi^2)$, positive lengths are once again only assigned to vectors along the spherical coordinates, not $v$. In contrast to $\Omega_S$, any cross section $\Sigma\hookrightarrow\mathcal{H}$ has metric $\gamma = 4M^2\mathring\gamma$. Thus, all cross-sections exhibit the exact same area $16\pi M^2$. In other words, on any cross-section of $\mathcal{H}$, light rays emitted perpendicularly off of the surface remain trapped. Since no material particles travel faster than the speed of light, $\mathcal{H}$ indicates the hypersurface from which there is `no return' upon entering, or the \textit{event horizon}. As a result of this trapping, the 2-sphere $\mathcal{H}\cap\Omega_S$ is the unique cross-section of $\Omega_S$ satisfying $\langle\vec{H},\vec{H}\rangle = 0$, namely with null mean curvature.
\begin{definition}
\emph{A \textit{marginally outer trapped surface} (MOTS) $\Sigma_0$ in an expanding null cone $\Omega$ is a surface satisfying the condition}
$$\langle\vec{H},\vec{H}\rangle = 0.$$
\end{definition}
\noindent With a MOTS $\Sigma_0$ identifying the presence of a black hole horizon in our context, the Null Penrose Conjecture states that
$$\sqrt{\frac{|\Sigma_0|}{16\pi}}\leq m_B.$$

\begin{figure}[H]
\centering
\def\svgwidth{\linewidth}
\fbox{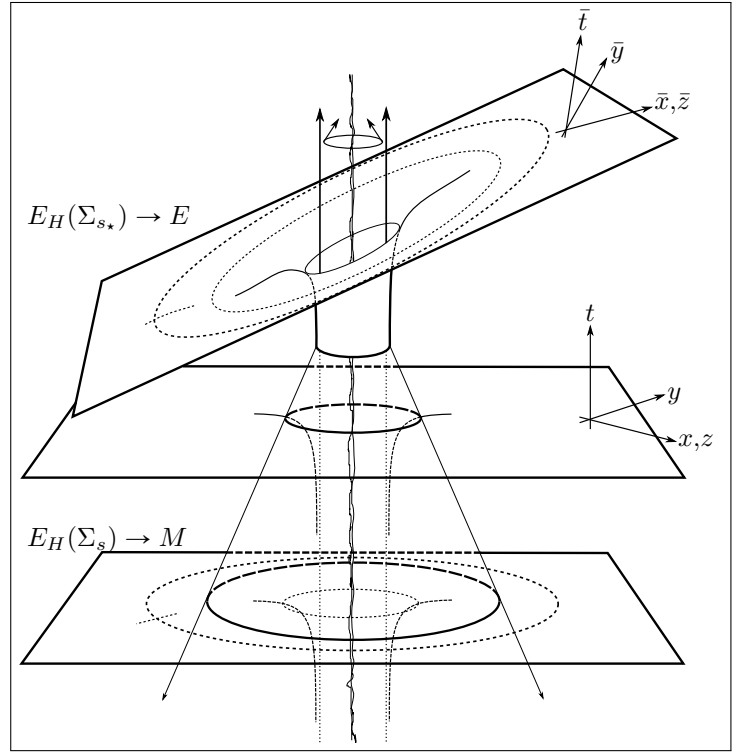}
\caption{Asymptotically round foliations of $\Omega_S$}
\end{figure}

\noindent\textbf{Previous Work}\\
Given a spacelike 2-sphere $\Sigma$ with metric $\gamma$, every point on its surface has two positive dimensions in the available four of spacetime spent on tangent vectors. As a result, we can combine the remaining negative and positive dimensions to form a normal null basis $\{\ubar L,L\}$. For a cross-section of a null cone $\Omega$ we choose $\ubar L$ as the normal-tangent to $\Omega$ (for example, see Figure 3). This finally allows us to introduce some final data for our main Theorem below.
\begin{definition}
\emph{For a smooth spacelike 2-sphere $\Sigma$ of second fundamental form $\II$ and null basis $\{\ubar L, L\}$ such that $\langle \ubar L,L\rangle = 2$ we define
$$\chi^-:= \langle -\II,\ubar L/\sigma\rangle,\,\,\zeta(V):=\frac12\langle D_V\ubar L,L\rangle=\tau(V)+V\log\sigma$$
where $\zeta$ is the \textit{connection 1-form} ($V$ a tangent vector field of $\Sigma$).}
\end{definition}
In his PhD thesis, Johannes Sauter (\cite{sauter2008foliations}) showed for the special class of \textit{shear free} null cones (i.e. satisfying $\chi^- = \frac12\gamma$) inside vacuum spacetimes, one is able to solve a system of ODEs to yield explicitly the geometry of $\Omega$. This then enables a direct analysis of $E_H$ at null infinity that allowed Sauter to prove the NPC. An observation of Christodoulou (see \cite{sauter2008foliations}) also shows that $E_H$ is monotonically increasing along foliations in vacuum if either the \textit{mass aspect function} $\mu: = \mathcal{K}-\frac14\langle\vec{H},\vec{H}\rangle-\nabla\cdot\zeta$ or the expansion $\sigma$ remain constant functions on each cross-section. For a black hole horizon $\Sigma_0$, we see from (1) that $E_H(\Sigma_0) = \sqrt{|\Sigma_0|/16\pi}$ making this observation particularly interesting. If one is successful in interpolating $E_H$ from the horizon to the Bondi Mass $m_B$ along one of these flows the NPC would follow for vacuum spacetimes $\sqrt{|\Sigma_0|/16\pi}=E_H(\Sigma_0)\leq \displaystyle{\lim_{s\to\infty}}m(\Sigma_s) = m_B$. Sauter was able to show for small pertubations of $\Omega$ off of the shear free condition, one obtains global existence of either of these flows and that $E_H$ converges. Unfortunately, one is unable to conclude that the foliating 2-spheres even become round asymptotically let alone $E_H$ approaching $m_B$. In fact, Bergqvist (\cite{bergqvist1997penrose}) noticed this exact difficulty had been overlooked in an earlier work of Ludvigsen and Vickers (\cite{ludvigsen1983inequality}) towards proving the \textit{weak NPC}, namely $\sqrt{|\Sigma_0|/16\pi}\leq E_B$.\\
\indent In 2015, Alexakis (\cite{alexakis2015penrose}) was able to prove the NPC for vacuum perturbations of the black hole exterior in Schwarzschild spacetime by successfully using the latter of the two flows in Sauter's thesis. Alexakis was once again afforded an explicit analysis of $E_H$ at null infinity. Work by Mars and Soria (\cite{mars2015asymptotic}) followed soon afterwards in identifying the asymptotically flat condition on $\Omega$ to maintain an explicit limit of $\displaystyle{\lim_{s\to\infty}E_H(\Sigma_s)}$ along geodesic foliations. In 2016 (\cite{mars2016penrose}), those authors constructed a new functional on 2-spheres and showed for a special foliation $\{\Sigma_\lambda\}$ off of the horizon $\Sigma_0$ called \textit{geodesic asymptotically Bondi} (or GAB) that, $\sqrt{|\Sigma_0|/16\pi}\leq\displaystyle{\lim_{\lambda\to\infty}E_H(\Sigma_\lambda)}<\infty$. Thus, for GAB foliations that approach round spheres, this reproduces the weak NPC of Bergqvist (\cite{bergqvist1997penrose}), Ludvigsen and Vickers (\cite{ludvigsen1983inequality}). Unfortunately, as in the aforementioned work of Sauter, Bergqvist, Ludvigsen and Vickers there is no guarantee of asymptotic roundness.\\\\
\textbf{Mass Not Energy}\\
These difficulties may very well be symptomatic of the fact that an energy is particularly susceptible to the plethora of ways boosts can develop along any given flow.\\
\indent We expect an infinitesimal null flow of $\Sigma$ to gain energy due to an influx of matter analogous to the addition of 4-velocities in Figure 5, $E_3 = E_1+E_2$. However, with energy being a frame dependent measurement and no way to discern a reference frame, we are left at the mercy of distortions along the flow. Without knowing, our measurements could experience either an artificial increase in energy $P\to P'$  or decrease $P'\to P$, as depicted in Figure 5. Geometrically, this manifests along the flow in a (local) `tilting' of $\Sigma$ (recall Figure 4). From the formula of $E_H(\Sigma)$ for $\Sigma\hookrightarrow\Omega_S$, Jensen's inequality indicates the existence of many flows with increasing $E_H$ yet only $t$-slice intersections produce the Bondi Mass. Not only is this flow highly specialized, it dictates strong restrictions on our initial choice of $\Sigma$ from which to begin the flow.\\

\begin{figure}[H]
\centering
\def\svgwidth{\linewidth} 
\fbox{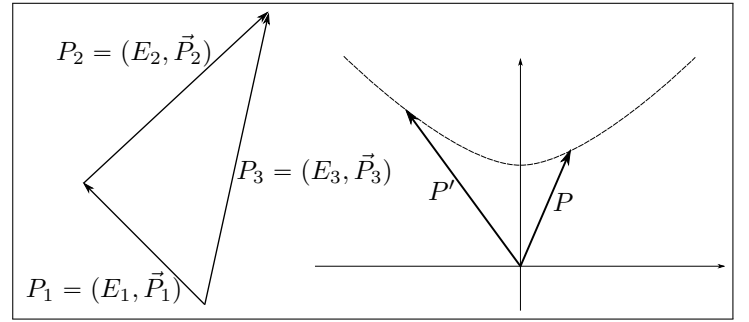}
\caption{Propagation vs Boosts}
\end{figure}

This is not a problem, however, if appealing instead to mass rather than energy since boosts leave mass invariant, $M^2=E^2-|\vec{P}|^2= (E')^2-|\vec{P}'|^2=(M')^2$. Moreover, by virtue of the Lorentzian triangle inequality (provided all vectors are timelike and either all future or all past-pointing), along any given flow the mass should always increase
$$M_3 = |(E_1+E_2,\vec{P_1}+\vec{P_2})|\geq |(E_1,\vec{P}_1)|+|(E_2,\vec{P}_2)| = M_1+M_2.$$
One may hope therefore, by appealing to a notion of mass instead of energy, a larger class of flows will arise exhibiting monotonicity of mass and physically meaningful asymptotics. \\\\
\noindent\textbf{Recent Progress from a new Quasi-local Mass}\\
In search of a mass we return to our favorite model spacetime, the Schwarzschild spacetime with \textit{null cones} $\Omega_S$. With the tantalizingly simple expression (3), a natural first guess at extracting the black hole mass $M$ is to take 
$$\tilde{m}(\Sigma)=\frac12\Big(\frac{1}{4\pi}\int(\mathcal{K}-\frac14\langle\vec{H},\vec{H}\rangle)^{\frac23} dA\Big)^{\frac32}.$$ 
The reason being, irrespective of the cross-section $\Sigma\hookrightarrow\Omega_S$, this gives $\tilde{m}(\Sigma) = \frac12\Big(\frac{1}{4\pi}\int(\frac{2M}{\omega^3})^{\frac23}(\omega^2dS)\Big)^{\frac32}=M$ as desired. Amazingly, Jensen's inequality also ensures that $\tilde{m}\leq E_H$ whenever the integrand is non-negative by Gauss-Bonnet. Unfortunately, upon an analysis of the propagation of this mass on a general null cone, no clear monotonicity properties arise and we're left needing to modify it at the very least. However, on a return to our drawing board $\Omega_S$, one finds that all cross-sections also satisfy $\zeta=d\log\sigma\leftrightarrow \tau = 0$ and a sufficient condition \cite{wang2014minkowski}. Inspired by this, the second author (see \cite{roesch2016proof}) put forward the modified \textit{geometric flux}, $\rho$ and mass $m(\Sigma)$:
\begin{align*}
&\rho:=\mathcal{K}-\frac14\langle\vec{H},\vec{H}\rangle+\nabla\cdot\zeta-\Delta\log\sigma\\
&m(\Sigma):=\frac12\Big(\frac{1}{4\pi}\int\rho^{\frac23}dA\Big)^{\frac32}.
\end{align*}
The first thing we observe, is that the previously desired properties $m(\Sigma) = M$ for $\Sigma\hookrightarrow\Omega_S$ and $m\leq E_H$ if $\rho\geq0$ are maintained (the second property now following from both the Guass-Bonnet and the Divergence Theorems). Even better, from a nine page calculation followed by three different `integrations by parts' most terms combine to ensure this mass function is nondecreasing in great generality, as summarized by our main theorem.\begin{theorem}(\cite{roesch2016proof})
Let $\Omega$ be a null cone foliated by spacelike spheres $\{\Sigma_s\}$ expanding along a null flow vector $\ubar L = \sigma L^-$ such that $|\rho(s)|>0$ for each $s$. Then the mass $m(s):=m(\Sigma_s)$ has rate of change
\begin{align*}
\frac{dm}{ds}=&\\\frac{(2m)^{\frac13}}{8\pi}&\int_{\Sigma_s}\frac{\sigma}{\rho^{\frac13}}\Big(|\hat{\chi}^-|^2+G(L^-,L^-))(\frac14\langle\vec{H},\vec{H}\rangle-\frac13\Delta\log|\rho|)\\
&\hspace{1.6in}+\frac12|\nu|^2+G(L^-,N)\Big)dA,
\end{align*}
where\\
$G=Ric-\frac12 Rg$ the Einstein tensor, $\nu=\frac23\hat{\chi}^-\cdot d\log|\rho|-\tau$, and $N = \frac19|d\log|\rho||^2L^-+\frac13\nabla\log|\rho|-\frac14L^+$.
\end{theorem}
From the DEC it follows that $G(L^-,L^-), G(L^-,N)\geq 0$, so that the convexity conditions $\rho>0$, $\frac14\langle\vec{H},\vec{H}\rangle\geq\frac13\Delta\log\rho$ along a foliation $\{\Sigma_s\}$, called a \textit{doubly convex foliation}, implies 
$$\boxed{\frac{dm}{ds}\geq 0.}$$ 
So how likely is a foliation to be doubly convex? Well, in the case of a cross-section of $\Omega_S$ in Schwarzschild, we know (4) holds trivially from (3). It also follows (\cite{roesch2016proof}) that $\frac14\langle\vec{H},\vec{H}\rangle-\frac13\Delta\log\rho = \frac{1}{\omega^2}(1-\frac{2M}{\omega})$. We conclude that all foliations of $\Omega_S$ in the black hole exterior ($\omega\geq 2M$) satisfy (4) and (5). A natural question follows as to whether these conditions are physically motivated for more general asymptotically flat null cones.\\
\indent Having found that this mass functional exhibits somewhat generic monotonicity, our next concern is asymptotic convergence and whether we obtain a physically significant quantity. From the fact that $m\leq E_H$, we see that any doubly convex foliation $\{\Sigma_s\}$ approaching a geodesic foliation of an asymptotically flat null cone $\Omega$ yields a converging mass, since $E_H(\Sigma_s)$ converges  (\cite{mars2015asymptotic}). However, this convergence is an indirect observation insufficient for a direct analysis of the limit. With a fairly standard strengthening of the decay conditions on the geometry of $\Omega$, called \textit{strong flux decay} (see \cite{roesch2016proof}) we're afforded an explicit limit for $\displaystyle{\lim_{s\to\infty}}m(\Sigma_s)$. Amazingly, this limit is independent of any choice of asymptotically geodesic foliation (as in Schwarzschild), and we conclude with a proof of a Null Penrose Conjecture. 
\begin{theorem}(\cite{roesch2016proof}) 
Let $\Omega$ be an asymptotically flat null cone with strong flux decay in a spacetime satisfying the DEC. Given the existence of an asymptotically geodesic doubly convex foliation $\{\Sigma_s\}$ 
$$\lim_{s\to\infty}m(\Sigma_s)\leq m_B.$$
Moreover, in the case that $\langle\vec{H},\vec{H}\rangle|_{\Sigma_0}=0$ (i.e. $\Sigma_0$ is a horizon) we prove the NPC. Furthermore, if we have the case of equality for the NPC and $\{\Sigma_s\}$ is a strict doubly convex foliation, then $\Omega=\Omega_S$.
\end{theorem}
The first part of the theorem follows from the fact that $m\leq E_H$ and that the limit $\displaystyle{\lim_{s\to\infty}}m(\Sigma_s)$ (being independent of the flow) must therefore bound all Bondi energies from below, hence also $m_B$. In the second part, if $\langle\vec{H},\vec{H}\rangle = 0$ along with (5), then its a consequence of the Maximum Principle for elliptic PDE that $\rho$ must be constant on $\Sigma_0$ warranting equality in Jensen's inequality. As a result, $$\sqrt{\frac{|\Sigma_0|}{16\pi}}=E_H(\Sigma_0)=m(\Sigma_0)\leq\displaystyle{\lim_{s\to\infty}}m(\Sigma_s)\leq m_B.$$ For the case of equality we refer the reader to \cite{roesch2016proof}.\\\\
\textbf{Open Problems}\\
An interesting condition resulting in a doubly convex foliation is that $\rho(s)$ be constant on each of the leaves of a foliation (i.e. on each $\Sigma_s$). As a result, this foliation satisfies $m(\Sigma_s) = E_H(\Sigma_s)$, representing a `rest-frame' flow given that energy equals mass. Studying the existence of this flow is of great interest.\\
\indent We also invite the reader to recall the dependence of $\zeta$ and $\sigma$ on the null basis $\{\ubar L,L\}$. An analogous construction of data under a `role reversal' between $\ubar L$ and $L$ would result in a new flux function $\bar{\rho}$ on our surface $\Sigma$. We say a surface $\tilde\Sigma$ is \textit{time-flat} (\cite{bray2014time}) whenever $\rho = \bar{\rho}$. The existence of `time-flat' surfaces within asymptotically flat null cones are particularly interesting as they serve as pivots where a flow can `flip-direction' without causing a discontinuous jump in mass (since $m = \bar{m}$).  

\begin{figure}[H]
\centering
\def\svgwidth{\linewidth} 
\fbox{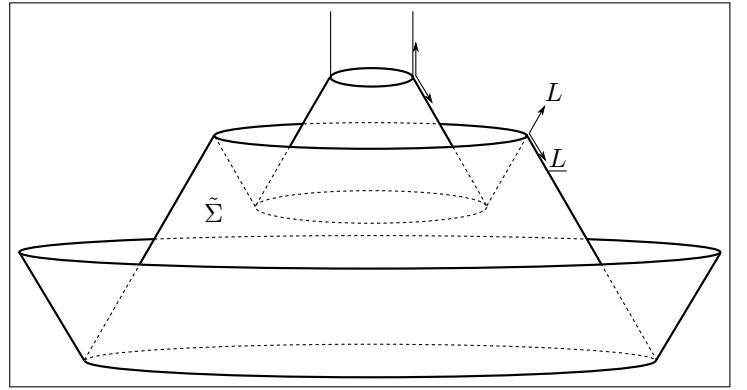}
\caption{Bouncing}
\end{figure}

This observation is of particular significance in the search for more general foliations that inherit the successes of Theorems 1 and 2. Another objective is to weaken the underlying smoothness assumption associated with null cones in Theorem 2, possibly toward broadening its validity to include focal points or multiple black hole horizons.\\
\indent These open questions are important not only for understanding the physics of black holes and mass in general relativity, but also for expanding our knowledge of null geometry. Since null geometry is not particularly intuitive, physical motivations like these are very useful for providing fascinating conjectures to pursue.

\bibliographystyle{abbrv}
\bibliography{Notices_ArXiv.bbl}
  \end{multicols}
\end{document}

%% file: SpecialR.pdf_tex
\begingroup%
  \makeatletter%
  \providecommand\color[2][]{%
    \errmessage{(Inkscape) Color is used for the text in Inkscape, but the package 'color.sty' is not loaded}%
    \renewcommand\color[2][]{}%
  }%
  \providecommand\transparent[1]{%
    \errmessage{(Inkscape) Transparency is used (non-zero) for the text in Inkscape, but the package 'transparent.sty' is not loaded}%
    \renewcommand\transparent[1]{}%
  }%
  \providecommand\rotatebox[2]{#2}%
  \ifx\svgwidth\undefined%
    \setlength{\unitlength}{775.93321312bp}%
    \ifx\svgscale\undefined%
      \relax%
    \else%
      \setlength{\unitlength}{\unitlength * \real{\svgscale}}%
    \fi%
  \else%
    \setlength{\unitlength}{\svgwidth}%
  \fi%
  \global\let\svgwidth\undefined%
  \global\let\svgscale\undefined%
  \makeatother%
  \begin{picture}(1,0.71115696)%
    \put(0,0){\includegraphics[width=\unitlength,page=1]{SpecialR.pdf}}%
    \put(0.05087703,0.49754604){\color[rgb]{0,0,0}\makebox(0,0)[lb]{\smash{}}}%
    \put(0.06025039,0.49025562){\color[rgb]{0,0,0}\makebox(0,0)[lb]{\smash{$t$}}}%
    \put(0.28937688,0.00284121){\color[rgb]{0,0,0}\makebox(0,0)[lb]{\smash{$x,z$}}}%
    \put(0.82261668,0.6876322){\color[rgb]{0,0,0}\makebox(0,0)[lb]{\smash{$\bar{t}$}}}%
    \put(0.86948345,0.63502192){\color[rgb]{0,0,0}\makebox(0,0)[lb]{\smash{$\bar{y}$}}}%
    \put(0.88302278,0.55170324){\color[rgb]{0,0,0}\makebox(0,0)[lb]{\smash{$\bar{x}$,$\bar{z}$}}}%
    \put(0.66877308,0.18986418){\color[rgb]{0,0,0}\makebox(0,0)[lb]{\smash{$\alpha$}}}%
    \put(0,0){\includegraphics[width=\unitlength,page=2]{SpecialR.pdf}}%
    \put(0.0520742,0.63634671){\color[rgb]{0,0,0}\makebox(0,0)[lb]{\smash{$\mathbb{R}^4_1$}}}%
  \end{picture}%
\endgroup%

%% file: cone1.pdf_tex
\begingroup%
  \makeatletter%
  \providecommand\color[2][]{%
    \errmessage{(Inkscape) Color is used for the text in Inkscape, but the package 'color.sty' is not loaded}%
    \renewcommand\color[2][]{}%
  }%
  \providecommand\transparent[1]{%
    \errmessage{(Inkscape) Transparency is used (non-zero) for the text in Inkscape, but the package 'transparent.sty' is not loaded}%
    \renewcommand\transparent[1]{}%
  }%
  \providecommand\rotatebox[2]{#2}%
  \ifx\svgwidth\undefined%
    \setlength{\unitlength}{488.40206116bp}%
    \ifx\svgscale\undefined%
      \relax%
    \else%
      \setlength{\unitlength}{\unitlength * \real{\svgscale}}%
    \fi%
  \else%
    \setlength{\unitlength}{\svgwidth}%
  \fi%
  \global\let\svgwidth\undefined%
  \global\let\svgscale\undefined%
  \makeatother%
  \begin{picture}(1,0.43799821)%
    \put(0,0){\includegraphics[width=\unitlength,page=1]{cone1.pdf}}%
    \put(0.68379887,0.39316251){\color[rgb]{0,0,0}\makebox(0,0)[lb]{\smash{$\Sigma_0$}}}%
    \put(0.82536825,0.14629327){\color[rgb]{0,0,0}\makebox(0,0)[lb]{\smash{$\Sigma_s$}}}%
    \put(0.03679087,0.36742251){\color[rgb]{0,0,0}\makebox(0,0)[lb]{\smash{$\Omega$}}}%
    \put(0,0){\includegraphics[width=\unitlength,page=2]{cone1.pdf}}%
    \put(0.30589008,0.30307283){\color[rgb]{0,0,0}\makebox(0,0)[lb]{\smash{$\ubar L$}}}%
    \put(0.59604908,0.31711276){\color[rgb]{0,0,0}\makebox(0,0)[lb]{\smash{$\ubar L$}}}%
    \put(0,0){\includegraphics[width=\unitlength,page=3]{cone1.pdf}}%
  \end{picture}%
\endgroup%

%% file: cone1a.pdf_tex
\begingroup%
  \makeatletter%
  \providecommand\color[2][]{%
    \errmessage{(Inkscape) Color is used for the text in Inkscape, but the package 'color.sty' is not loaded}%
    \renewcommand\color[2][]{}%
  }%
  \providecommand\transparent[1]{%
    \errmessage{(Inkscape) Transparency is used (non-zero) for the text in Inkscape, but the package 'transparent.sty' is not loaded}%
    \renewcommand\transparent[1]{}%
  }%
  \providecommand\rotatebox[2]{#2}%
  \ifx\svgwidth\undefined%
    \setlength{\unitlength}{364.65893843bp}%
    \ifx\svgscale\undefined%
      \relax%
    \else%
      \setlength{\unitlength}{\unitlength * \real{\svgscale}}%
    \fi%
  \else%
    \setlength{\unitlength}{\svgwidth}%
  \fi%
  \global\let\svgwidth\undefined%
  \global\let\svgscale\undefined%
  \makeatother%
  \begin{picture}(1,0.85242875)%
    \put(0,0){\includegraphics[width=\unitlength,page=1]{cone1a.pdf}}%
    \put(0.42434554,0.81753781){\color[rgb]{0,0,0}\makebox(0,0)[lb]{\smash{$r=0$}}}%
    \put(0.57131436,0.75639392){\color[rgb]{0,0,0}\makebox(0,0)[lb]{\smash{$r=2M$}}}%
    \put(0.14564607,0.15539647){\color[rgb]{0,0,0}\makebox(0,0)[lb]{\smash{$\omega=r|_\Sigma$}}}%
    \put(0.6868767,0.25186807){\color[rgb]{0,0,0}\makebox(0,0)[lb]{\smash{$L$}}}%
    \put(0.67968765,0.13573468){\color[rgb]{0,0,0}\makebox(0,0)[lb]{\smash{$\ubar L$}}}%
    \put(-0.24740048,0.27893395){\color[rgb]{0,0,0}\makebox(0,0)[lb]{\smash{}}}%
    \put(-0.03898657,0.33534675){\color[rgb]{0,0,0}\makebox(0,0)[lb]{\smash{}}}%
    \put(0.10578603,0.77672034){\color[rgb]{0,0,0}\makebox(0,0)[lb]{\smash{$\Omega_S$}}}%
    \put(0,0){\includegraphics[width=\unitlength,page=2]{cone1a.pdf}}%
    \put(0.82776464,0.24821366){\color[rgb]{0,0,0}\makebox(0,0)[lb]{\smash{$\Sigma$}}}%
    \put(0.94789524,0.65153842){\color[rgb]{0,0,0}\makebox(0,0)[lb]{\smash{}}}%
    \put(0,0){\includegraphics[width=\unitlength,page=3]{cone1a.pdf}}%
    \put(-0.04197264,0.99725059){\color[rgb]{0,0,0}\makebox(0,0)[lt]{\begin{minipage}{1.17896715\unitlength}\raggedright \end{minipage}}}%
  \end{picture}%
\endgroup%

%% file: slices.pdf_tex
\begingroup%
  \makeatletter%
  \providecommand\color[2][]{%
    \errmessage{(Inkscape) Color is used for the text in Inkscape, but the package 'color.sty' is not loaded}%
    \renewcommand\color[2][]{}%
  }%
  \providecommand\transparent[1]{%
    \errmessage{(Inkscape) Transparency is used (non-zero) for the text in Inkscape, but the package 'transparent.sty' is not loaded}%
    \renewcommand\transparent[1]{}%
  }%
  \providecommand\rotatebox[2]{#2}%
  \ifx\svgwidth\undefined%
    \setlength{\unitlength}{492.20086526bp}%
    \ifx\svgscale\undefined%
      \relax%
    \else%
      \setlength{\unitlength}{\unitlength * \real{\svgscale}}%
    \fi%
  \else%
    \setlength{\unitlength}{\svgwidth}%
  \fi%
  \global\let\svgwidth\undefined%
  \global\let\svgscale\undefined%
  \makeatother%
  \begin{picture}(1,1.0406543)%
    \put(-0.19271588,0.73408349){\color[rgb]{0,0,0}\makebox(0,0)[lb]{\smash{}}}%
    \put(0,0){\includegraphics[width=\unitlength,page=1]{slices.pdf}}%
    \put(0.79481686,1.01388456){\color[rgb]{0,0,0}\makebox(0,0)[lb]{\smash{$\bar{t}$}}}%
    \put(0.84060605,0.97416814){\color[rgb]{0,0,0}\makebox(0,0)[lb]{\smash{$\bar{y}$}}}%
    \put(0.90314352,0.9012098){\color[rgb]{0,0,0}\makebox(0,0)[lb]{\smash{$\bar{x}$,$\bar{z}$}}}%
    \put(0.80527982,0.59933925){\color[rgb]{0,0,0}\makebox(0,0)[lb]{\smash{$t$}}}%
    \put(0.92195718,0.48849239){\color[rgb]{0,0,0}\makebox(0,0)[lb]{\smash{$y$}}}%
    \put(0.9370633,0.42219447){\color[rgb]{0,0,0}\makebox(0,0)[lb]{\smash{$x$,$z$}}}%
    \put(0,0){\includegraphics[width=\unitlength,page=2]{slices.pdf}}%
    \put(0.01149294,0.73768518){\color[rgb]{0,0,0}\makebox(0,0)[lb]{\smash{$\tiny{E_H(\Sigma_{s_\star})\to E}$}}}%
    \put(0,0){\includegraphics[width=\unitlength,page=3]{slices.pdf}}%
    \put(0.01129034,0.27983373){\color[rgb]{0,0,0}\makebox(0,0)[lb]{\smash{$\tiny{E_H(\Sigma_s)\to M}$}}}%
    \put(0,0){\includegraphics[width=\unitlength,page=4]{slices.pdf}}%
  \end{picture}%
\endgroup%

%% file: figures.pdf_tex
\begingroup%
  \makeatletter%
  \providecommand\color[2][]{%
    \errmessage{(Inkscape) Color is used for the text in Inkscape, but the package 'color.sty' is not loaded}%
    \renewcommand\color[2][]{}%
  }%
  \providecommand\transparent[1]{%
    \errmessage{(Inkscape) Transparency is used (non-zero) for the text in Inkscape, but the package 'transparent.sty' is not loaded}%
    \renewcommand\transparent[1]{}%
  }%
  \providecommand\rotatebox[2]{#2}%
  \ifx\svgwidth\undefined%
    \setlength{\unitlength}{833.91109437bp}%
    \ifx\svgscale\undefined%
      \relax%
    \else%
      \setlength{\unitlength}{\unitlength * \real{\svgscale}}%
    \fi%
  \else%
    \setlength{\unitlength}{\svgwidth}%
  \fi%
  \global\let\svgwidth\undefined%
  \global\let\svgscale\undefined%
  \makeatother%
  \begin{picture}(1,0.42532268)%
    \put(0,0){\includegraphics[width=\unitlength,page=1]{figures.pdf}}%
    \put(0.04967303,0.35661159){\color[rgb]{0,0,0}\makebox(0,0)[lb]{\smash{$P_2=(E_2,\vec{P}_2)$}}}%
    \put(0.00672976,0.01964211){\color[rgb]{0,0,0}\makebox(0,0)[lb]{\smash{$P_1=(E_1,\vec{P}_1)$}}}%
    \put(0,0){\includegraphics[width=\unitlength,page=2]{figures.pdf}}%
    \put(0.75575039,0.14145445){\color[rgb]{0,0,0}\makebox(0,0)[lb]{\smash{$P$}}}%
    \put(0.57852935,0.15731377){\color[rgb]{0,0,0}\makebox(0,0)[lb]{\smash{$P'$}}}%
    \put(0.17379527,0.30092069){\color[rgb]{0,0,0}\makebox(0,0)[lb]{\smash{}}}%
    \put(0.30592822,0.18719635){\color[rgb]{0,0,0}\makebox(0,0)[lb]{\smash{$P_3=(E_3,\vec{P}_3)$}}}%
  \end{picture}%
\endgroup%

%% file: bounce.pdf_tex
\begingroup%
  \makeatletter%
  \providecommand\color[2][]{%
    \errmessage{(Inkscape) Color is used for the text in Inkscape, but the package 'color.sty' is not loaded}%
    \renewcommand\color[2][]{}%
  }%
  \providecommand\transparent[1]{%
    \errmessage{(Inkscape) Transparency is used (non-zero) for the text in Inkscape, but the package 'transparent.sty' is not loaded}%
    \renewcommand\transparent[1]{}%
  }%
  \providecommand\rotatebox[2]{#2}%
  \ifx\svgwidth\undefined%
    \setlength{\unitlength}{522.76889591bp}%
    \ifx\svgscale\undefined%
      \relax%
    \else%
      \setlength{\unitlength}{\unitlength * \real{\svgscale}}%
    \fi%
  \else%
    \setlength{\unitlength}{\svgwidth}%
  \fi%
  \global\let\svgwidth\undefined%
  \global\let\svgscale\undefined%
  \makeatother%
  \begin{picture}(1,0.52235623)%
    \put(0,0){\includegraphics[width=\unitlength,page=1]{bounce.pdf}}%
    \put(0.74896409,0.39419165){\color[rgb]{0,0,0}\makebox(0,0)[lb]{\smash{$L$}}}%
    \put(0.7543242,0.30037835){\color[rgb]{0,0,0}\makebox(0,0)[lb]{\smash{$\ubar L$}}}%
    \put(0.91359705,0.39651042){\color[rgb]{0,0,0}\makebox(0,0)[lb]{\smash{}}}%
    \put(0.26434022,0.22260228){\color[rgb]{0,0,0}\makebox(0,0)[lb]{\smash{$\tilde\Sigma$}}}%
  \end{picture}%
\endgroup%